\documentclass[twocolumn,showpacs,preprintnumbers,nofootinbib,amsmath,amssymb,superscriptaddress,aps,prc]{revtex4-1}
\usepackage{graphicx}% Include figure files
\usepackage{dcolumn}% Align table columns on decimal point
\usepackage{bm}% bold math
\usepackage{float}
\usepackage{caption}
\usepackage{subcaption}

\graphicspath {{figures/}}

\usepackage{hyperref}
\hypersetup{
     colorlinks   = true,
     linkcolor = cyan,
     citecolor = cyan,
     menucolor = black,
     urlcolor = cyan  
}

\usepackage{comment}

\begin{document}

\title{Microscopic predictions of the nuclear matter liquid-gas phase transition}% Force line breaks with \\
 
\author{Arianna Carbone}
\email[Email:~]{acarbone@ectstar.eu}
\affiliation{European Centre for Theoretical Studies in Nuclear Physics and Related Areas (ECT*)
and Fondazione Bruno Kessler, Strada delle Tabarelle 286, I-38123 Villazzano (TN), Italy}

\author{Artur Polls}
\email[Email:~]{artur@fqa.ub.edu}
\affiliation{Departament de F\'isica Qu\`antica i Astrof\'isica  and Institut de 
Ci\`{e}ncies del Cosmos (ICCUB), Universitat de Barcelona, E-08028 Barcelona, Spain}
 
\author{Arnau Rios}
\email[Email:~]{a.rios@surrey.ac.uk}
\affiliation{Department of Physics, Faculty of Engineering and Physical Sciences, University of Surrey, Guildford, Surrey GU2 7XH, United Kingdom}

\date{\today}% It is always \today, today,

\begin{abstract}
We present first-principle predictions for the liquid-gas phase transition in symmetric nuclear matter employing both two- and three-nucleon chiral interactions. Our discussion focuses on the sources of systematic errors in microscopic quantum many body predictions. On the one hand, we test uncertainties of our results arising from changes in the construction of chiral Hamiltonians. We use five different chiral forces with consistently derived three-nucleon interactions. On the other hand, we compare the ladder resummation in the self-consistent Green's functions approach to finite temperature Brueckner--Hartree--Fock calculations. We find that systematics due to Hamiltonians dominate over many-body uncertainties. 
Based on this wide pool of calculations, we estimate that the critical temperature is $T_c=16 \pm 2$ MeV, 
in reasonable agreement with experimental results. We also find that there is a strong correlation between the critical temperature and the saturation energy in microscopic many-body simulations. 
\end{abstract}

\maketitle

\section{Introduction}
\iffalse
\begin{itemize}
\item The liquid-gas phase transition in nuclear matter
 \item Experimental evidence
 \item Past theoretical calculations
 \item Present calculations
 \item Stuff in the paper
\end{itemize}
\fi

The quest for understanding the phase diagram of strongly-interacting hadronic matter is a contemporary physics challenge, both from an experimental and a theoretical perspective \cite{BraunMunziger2009,Fukushima2011,Halasz1998}. Recent advances have focused on the deconfinement phase transition, but a complete picture is hard to obtain due to, among other things, the complexity of the strong interaction as given by quantum chromodynamics (QCD).  At comparatively lower temperatures and densities, the liquid-gas phase transition of nuclear matter is an equally challenging phenomenon~\cite{Siemens1983}. This phase transition is expected to exist on the grounds that the nucleon-nucleon and the inter-atomic Van der Waals interactions are qualitatively similar~\cite{Fisher1967}. The first-order liquid-gas phase transition has been investigated experimentally in the past, via multifragmentation reactions and through the analysis of fission fragments emitted by excited compound nuclei~\cite{Finn1982,Pochodzalla1997,Karnaukhov2009,Elliott2013}. A key indication of an actual phase transition is the power law of the fragment distributions in both types of reactions. When extrapolated to infinite matter, the critical temperature $T_c$ estimated with these finite-nucleus experiments lies in a range $T_c \approx 15-20$ MeV. 

In spite of the experimental signatures of its existence, a microscopic modeling of this phase transition faces stringent difficulties. On the one hand, the nucleon-nucleon force is not entirely well understood, and cannot be univocally derived from QCD itself \cite{Epelbaum2009,Machleidt2011}. On the other, the phase transition requires solving the problem of strongly-interacting many-body fermions at finite temperature \cite{Fetter2003}. A first-principles understanding of the liquid-gas transition in nuclear matter is necessary to connect unambiguously these experimental results to the strong interaction. 

Moreover, the tools that are necessary to model microscopically this phenomenon directly relate to several aspects of astrophysical interest. Finite temperature effects are relevant for the evolution of protoneutron stars \cite{Prakash1997}. In a binary merger, the gravitational wave spectrum, neutrino emission and ejecta distribution are sensitive to the equation of state which, in turn, is temperature-dependent \cite{Sekiguchi2011,Radice2016}. The cooling process of neutron stars via neutrino emission is influenced by temperature-sensitive in-medium effects \cite{Rrapaj2015}. A consistent picture describing all these aspects from a microscopic many-body perspective is still missing. 

A wide range of theoretical tools have been used to model the liquid-gas phase transition in the last four decades. There is no clear separation of scales in terms of densities or temperatures and one therefore needs to solve the problem head on using numerical techniques.  In doing so, there are a number of approximations that bring in systematic uncertainties in the predictions of the phase transition. It is these systematics that we want to analyze in this paper. Early phenomenological models exploited the idea of nuclear condensation to understand the appearance of critical fragment distributions~\cite{Jaqaman1983,Kapusta1984}. Self-consistent Hartree-Fock or density functional ideas have also been used to describe this transition with different effective interactions~\cite{Sauer1976,Lamb1981,BALi2008,Rios2010,Carbone2011,Lourenco2016}. Beyond-mean-field calculations have relied on phase-shift-equivalent two-nucleon (2N) interactions as a starting point~\cite{Friedman1981,Baldo1999,Rios2008,Soma2009}. We will focus on the latter kind of microscopic many-body models that treat infinite nuclear matter, without explicitly considering fragment production or clustering. We will solve the phase transition using a Maxwell construction of two infinite, coexisting phases, using two different many-body approaches. By looking at the problem with two different many-body techniques, we expect to highlight systematic errors associated to the many-body approximation. 

Another source of uncertainty in theoretical calculations stems from the nuclear Hamiltonian itself. In contrast to traditional microscopic phase-shift equivalent 2N forces, chiral effective field theory (EFT) provides a systematic grouping of nuclear forces in terms of a power counting expansion. An important advantage is that the expansion also provides a clear organization in terms of two- and many-nucleon contributions~\cite{Epelbaum2009,Machleidt2011} and allows for statistical error quantification \cite{Carlsson2016}. We will therefore use chiral EFT 2N and three-nucleon (3N) forces to describe the phase transition. Recently, the liquid-gas transition in nuclear matter has been studied within many-body perturbation theory based on such chiral interactions~\cite{Wellenhofer2014,Wellenhofer2015}. We go beyond these past studies in two different ways. First, we estimate errors in the many-body calculation by using two different first-principles schemes to solve the (finite-temperature) many-body problem. Second, past quantum many-body calculations have not necessarily explored the systematics associated with the use of different 2N and 3N forces. Here, we aim at improving over these limitations by making use of recent nuclear forces derived consistently within chiral EFT. We estimate uncertainties coming from the chiral Hamiltonian by employing different 2N forces, including the recently developed N2LO$_{\rm sat}$ (2N+3N) potential~\cite{Ekstroem2015}. 

The paper is organized as follows. Section~\ref{sec:formalism} is divided into three parts. In the first two parts, Secs.~\ref{subsec:scgf_form} and \ref{finiteT_BHF}, we describe respectively the self-consistent Green's function (SCGF) and the Brueckner--Hartree--Fock (BHF) formalisms that we employ to study the liquid-gas phase transition. In the third part, Sec.~\ref{subsec:chiral_ham}, we provide details of the Hamiltonians used in our calculations. We then follow with a section of results, Sec.~\ref{sec:results}, which is itself divided into four parts. Section~\ref{subsec:therm_quant} reports on the free energy, chemical potential and pressure at different densities and temperatures, using different chiral Hamiltonians and analyzing the errors related to the many-body approximation. Section~\ref{subsec:liquid-gas} analyzes the characteristics of the liquid-gas phase transition for a single Hamiltonian: the N2LO$_{\rm sat}$ potential. In Sec. \ref{subsec:coex_line} we extend this analysis employing five different chiral forces and estimating the uncertainty on the critical temperature. The last part, Sec.~\ref{subsec:crit_sat}, is dedicated to the prediction of the critical temperature in relation to other properties of nuclear matter. We drive some conclusions on this study in the final section.

\section{Formalism}
\label{sec:formalism}
\subsection{Finite-temperature Green's functions}
\label{subsec:scgf_form}
\iffalse
\begin{itemize}
\item Finite-T SCGF: free energy, pressure, chemical potential
\item thermodynamical consistency
\item evaluation of the liquid-gas phase transition
\end{itemize}
\fi

The SCGF is particularly suited for the study of finite temperature many-body dynamics because it provides, in principle, a thermodynamically consistent description~\cite{Baym1961,Baym1962}. In other words, the many-body description is conserving so that microscopic and macroscopic estimates of physical properties provide the same results. This is possible thanks to the Luttinger-Ward expression for the partition function in terms of the dressed one-body propagator, i.e. the Green's function~\cite{Kohn1960,Luttinger1960II}. 
Combining the Luttinger-Ward formalism with an extended SCGF method to include three-body forces~\cite{Carbone2014}, we predict the critical behavior of the first-order phase transition in symmetric nuclear matter considering both 2N and 3N forces. We note that, within the \emph{ab initio} SCGF method, results of the liquid-gas transition have already been presented in the past using both 2N interactions~\cite{Rios2008} and also Urbana-type 3N forces with a crude averaging procedure~\cite{Soma2009}. 

\begin{figure}[t!]
	\includegraphics[scale=0.55]{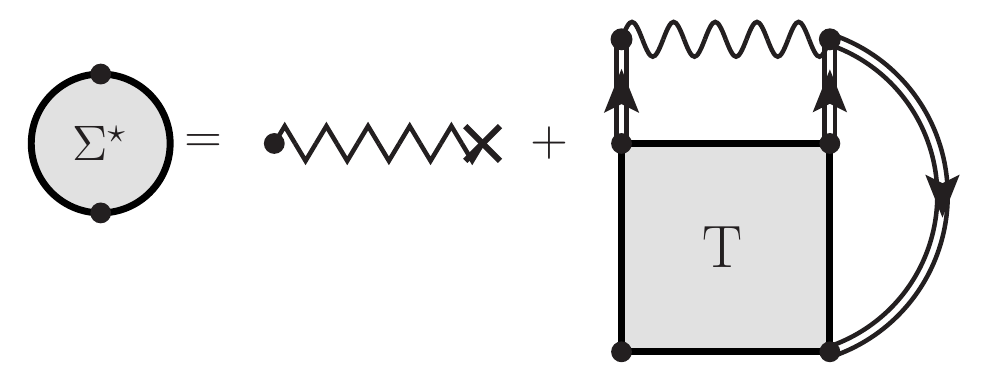}
	\caption{Diagrammatic representation of the irreducible self-energy. The first term represents the energy-independent one-body effective interaction depicted in Fig.~\ref{subfig:ueff}. The second term is a dispersive contribution arising from the two-body effective interaction in Fig.~\ref{subfig:veff}. The box in this term represents the in-medium interaction or $T$-matrix.}
\label{fig:self_energy}
\end{figure}

In the SCGF method, one looks for a self-consistent fully-correlated solution of the dressed single-particle Green's function at finite temperature. This method is based on the generalization to finite temperatures of the perturbation expansion of the one-body propagator $G$, using the Matsubara technique for the evaluation of imaginary frequency sums~\cite{Fetter2003}.  Knowledge of $G$ gives access to many properties of the system, such as the single-particle momentum distribution or the total energy per nucleon. We work with a Hamiltonian that is the sum of a free, unperturbed part, plus an interacting part, i.e. $\hat H=\hat H_0+\hat H_1$. Self-consistency in the Green's functions method is encoded in the Dyson equation, relating the unperturbed propagator, $G_0$, to the dressed single-particle propagator, $G$,
\begin{equation}
G({\bf p},\omega)=G_{0}({\bf p},\omega)+G_{0}({\bf p},\omega)\Sigma^\star({\bf p},\omega)G({\bf p},\omega)\,.
\label{eq:dyson}
\end{equation}   
${\bf p}$ and $\omega$ are the energy and momentum of single-particle states, which are good quantum numbers in infinite matter. $\Sigma^\star$ is the self-energy, which accounts for the interaction of a particle with the rest of the system. The self-energy can be expressed in terms of diagrams, as shown in Fig.~\ref{fig:self_energy}. 

In self-consistent calculations, only irreducible and skeleton diagrams are included. Diagrams are called irreducible if no diagram can be disconnected into two independent ones cutting a fermion line, and are said to be skeleton diagrams if no self-energy insertions are present~\cite{Carbone2013II,Mattuck}. 
All reducible diagrams are obtained via the Dyson equation, Eq.~(\ref{eq:dyson}), while self-energy insertions are taken into account by replacing all self-energy fermion lines with fully-dressed propagators, $G$. It is this self-consistent procedure that makes the approach thermodynamically consistent~\cite{Baym1961,Baym1962}.

We implement an in-medium ladder resummation scheme within the SCGF approach \cite{dickhoff04}. This non-perturbative approximation provides a realistic description of short-range correlations. Intermediate states in the in-medium $T$-matrix interaction are fully fragmented and include particle-particle and hole-hole states \cite{frick03,riosphd}.The diagrams in Fig.~\ref{fig:self_energy} illustrate the self-energy used in this work. The first term is a Hartree-Fock-like contribution, whereas the second term is a dispersive term associated to the in-medium $T$-matrix. This interaction is obtained from the solution of a Lippman-Schwinger-type equation \cite{Dickhoff08}.

\begin{figure}[t!]
   \centering
    
    \begin{subfigure}[t]{0.45\textwidth}
       \includegraphics[width=\textwidth]{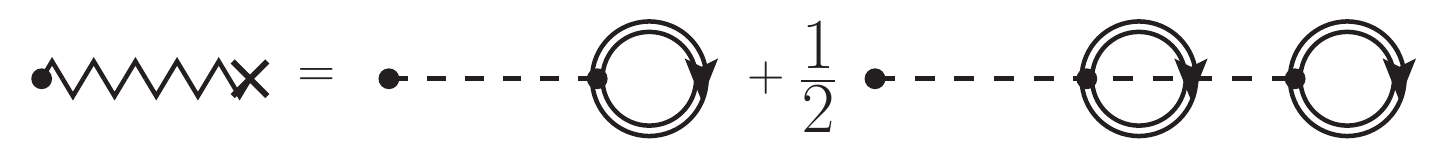}
       \caption{Diagrammatic representation of the one-body effective interaction. This is the sum of a 2N force contracted with a one-body propagator, plus a 3N force averaged with the lowest order two-body propagator. The factor $1/2$ defines the multiplicity of the contribution.}
       \label{subfig:ueff}
     \end{subfigure}

    \begin{subfigure}[t]{0.45\textwidth}
        \includegraphics[width=\textwidth]{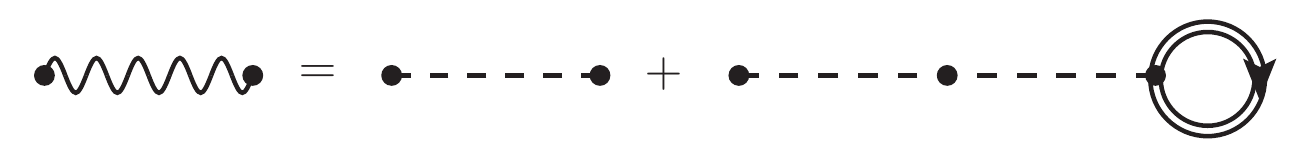}
        \caption{Diagrammatic representation of the two-body effective interaction. This is the sum of a bare 2N force plus a 3N force averaged with a one-body propagator.}
        \label{subfig:veff}
    \end{subfigure}
       
    \caption{Diagrams that define the (a) one-body and (b) two-body effective interactions in our approach.}
    \label{fig:eff_int}
\end{figure}

When the interacting Hamiltonian $\hat H_1=\hat V+ \hat W$ includes 2N forces, $\hat V$, and 3N forces, $\hat W$, one needs to pay particular attention to the correct definition of the self-energy $\Sigma^\star$~\cite{Carbone2013I}. The SCGF method has been generalized to include many-body interactions in Ref.~\cite{Carbone2013II}. 
To account for 3N forces when studying infinite nuclear matter within the SCGF method, we define effective one- and two-body interactions which are obtained calculating averages of the 3N force over dressed correlated single-particle propagators~\cite{Carbone2014}. These go beyond standard normal-ordering approaches by considering correlations and single-particle fragmentation on the averaging procedure. The diagrams associated to effective interactions are depicted in Fig.~\ref{subfig:ueff} and Fig.~\ref{subfig:veff}. The one-body effective interaction of Fig.~\ref{subfig:ueff} is a sum of a 2N force averaged over a one-body propagator plus a 3N force averaged over two one-body propagators. We note that, in the most general case, this average should be performed with a two-body propagator~\cite{Carbone2013II}. This one-body effective interaction is basically the energy-independent Hartree-Fock term of the self-energy. The second term, Fig.~\ref{subfig:veff}, is a sum of the bare 2N interaction plus a one-body averaged 3N force. This enters explicitly the definition of the in-medium $T$-matrix. For full formal expressions and discussions on the numerics, we refer the reader to Refs.~\cite{Carbone2014} and~\cite{Barbieri2017}. 

For a given temperature and density, we solve iteratively a self-consistent set of equations for the in-medium interaction and the self-energy. The key intermediate step is the calculation of the dressed single-particle propagator by means of the so-called spectral function. We can in fact recast the Dyson equation, Eq.~(\ref{eq:dyson}), as an expression for the spectral function,
\begin{equation}
A({\bf p},\omega)=\frac{\Gamma({\bf p},\omega)}{\big[\omega-\frac{p^2}{2m}-{\rm Re}\Sigma^\star({\bf p},\omega)\big]^2+\big[\frac{\Gamma({\bf p},\omega)}{2}\big]^2}\,,
\end{equation}
where $\Gamma({\bf p},\omega)$ is the imaginary part of the self-energy, $\Gamma({\bf p},\omega)=-2{\rm Im}\Sigma^\star({\bf p},\omega)$. For an energy-independent self-energy, the spectral function is a Lorentzian distribution of width $\Gamma$. In fact, the spectral function describes the probability of adding or removing a particle with momentum ${\bf p}$ and energy $\omega$ to or from the many-body system with a change in energy $d \omega$ \cite{Dickhoff08}. Consequently, for each momentum, $\Gamma$ is a proxy for the spread in energy of the single-particle strength. A vanishing $\Gamma$ results in a delta-like spectral function that provides a one-to-one quasi-particle relation between momentum and energy. The BHF approach is based on such a quasi-particle description \cite{zuo06}.  

Our self-consistent procedure provides fully fragmented spectral functions, $A({\bf p},\omega)$, for any given temperature and density. 
This function is then used to compute correlated temperature-dependent momentum distributions, which are in turn used to evaluate the effective interactions given in Fig.~\ref{fig:eff_int}. We note that the averaging procedure to evaluate the effective interactions of Fig.~\ref{subfig:ueff} and Fig.~\ref{subfig:veff} must be performed at each iterative step in the solution of the Dyson equation. In other words, we use propagators in the averaging which are the self-consistent solutions at that stage of the calculation. 

When the Hamiltonian is formed of 2N forces only, the Galitskii--Migdal--Koltun sum-rule provides access to the total energy of the system from the spectral function alone~\cite{Galitskii1958,Koltun1974,Dickhoff08}. When 3N forces are considered, the original sum-rule does not yield the total energy of the system and additional information from two-body or three-body operators is needed~\cite{Carbone2013II}. Here, we calculate the total energy per nucleon using the original sum-rule plus an estimate of the energy associated to the 3N force, 
\begin{equation}
\frac{E}{A}=\frac{4}{\rho}\int\frac{{\rm d}{\bf p}}{(2\pi)^3}\int\frac{{\rm d}\omega}{2\pi}\frac{1}{2}\Big\{\frac{p^2}{2m}+\omega\Big\}A({\bf p},\omega)f(\omega)-\frac{1}{2}\langle \hat W \rangle\,,
\label{eq:energy}
\end{equation}
where $f(\omega)$ is the Fermi-Dirac function and the factor of $4$ accounts for spin-isospin degeneracy. $\langle \hat W \rangle$ is the expectation value of the total energy associated to the 3N force. We calculate this at first order, i.e. by contracting $\hat W$ with three independent, but fully dressed momentum distributions. This is equivalent to a Hartree--Fock 3N force energy, where bare propagators are replaced by fully dressed ones~\cite{Carbone2013II}. 

While the formalism is defined in the grandcanonical ensemble, in practice we work at constant temperature and density in a canonical setting \cite{Fetter2003,frick03}. The thermodynamical properties are encoded in the free energy,
\begin{equation}
\frac{F}{A}=\frac{E}{A}-T \,\frac{S}{A}\,. 
\label{eq:free}
\end{equation}
The energy is provided by the modified sum-rule of Eq.~(\ref{eq:energy}).
To evaluate the entropy, $\frac{S}{A}$, we follow Luttinger and Ward, who demonstrated that it is possible to define the grand-canonical partition function $\Omega$ in terms of the dressed one-body propagator~\cite{Kohn1960,Luttinger1960II}. From this, given also the stationarity of $\Omega$ with respect to variations of single-particle propagators, $G$, one obtains the entropy via the derivative $S=-\frac{\partial \Omega}{\partial \rm T}|_\mu$. The pioneering work by Carneiro and Pethick showed that this entropy can be split into two terms, $S=S^\textrm{DQ}+S'$~\cite{Carneiro1975}. The first term defines a dynamical quasi-particle entropy, which takes into account the effects of correlations in the many-body system, and is given by the convolution:
\begin{equation}
\frac{S^\textrm{DQ}}{A}=\frac{4}{\rho} \int\frac{{\rm d}{\bf p}}{(2\pi)^3}\int\frac{\rm d\omega}{2\pi}\frac{\partial f(\omega)}{\partial\rm T}\Xi({\bf p},\omega)\,.
\label{eq:entropy_sq}
\end{equation}  
The quantity $\Xi$ is given by three terms:
\begin{align}
\Xi({\bf p},\omega)&= 2\pi\theta[{\rm Re}G^{-1}({\bf p},\omega)] \nonumber \\
& - 2 \arctan \left[ \frac{\Gamma({\bf p},\omega)}{2{\rm Re}G^{-1}({\bf p},\omega)} \right] \nonumber \\
& +\Gamma({\bf p},\omega){\rm Re}G({\bf p},\omega) \, .
\label{eq:xi_sq}
\end{align}
The first term provides a quasi particle-approximation to the entropy. The other two terms are zero whenever $\Gamma$ is zero \cite{rios06,riosphd}. They are therefore sensitive to fragmentation effects associated to correlations in the system.  
We work under the assumption that the term $S'$ can be disregarded due to phase space restrictions as suggested in Ref.~\cite{Carneiro1975}. 

Thermodynamical consistency is formally guaranteed in the ladder SCGF approach. Our numerical implementation can be tested by performing a comparison between two different determinations of the same physical quantity. For instance, the chemical potential can be obtained microscopically by inverting the density sum rule,
\begin{equation}
\rho=\frac{4}{\rho} \int\frac{{\rm d}{\bf p}}{(2\pi)^3}\int\frac{\rm d\omega}{2\pi}A({\bf p},\omega)f(\omega)\,.
\label{eq:mumicro}
\end{equation}
$\tilde\mu$ is the chemical potential in the Fermi-Dirac distribution function $f(\omega)=[1+e^{(\omega-\tilde\mu)/T}]^{-1}$. Alternatively, the chemical potential can also be determined macroscopically via the thermodynamical relation
\begin{equation}
\mu=\frac{\partial}{\partial\rho}\rho\frac{F}{A}\,,
\label{eq:mumacro}
\end{equation}
where $F/A$ is given in Eq.~\eqref{eq:free}. In a thermodynamically consistent theory, the equality $\tilde \mu=\mu$ holds \cite{riosphd}. In inconsistent approaches, such as BHF \cite{Baldo1999} or some implementations of many-body perturbation theory \cite{Wellenhofer2014}, the two determinations do not need to agree with each other.

The pressure of nuclear matter can also be determined from microscopic and macroscopic observables. The macroscopic pressure relies on the thermodynamical relation
\begin{equation}
P=\rho^2\frac{\partial F/A}{\partial\rho}\,.
\label{eq:pressmacro}
\end{equation}
In contrast, a microscopic calculation relies on the microscopic chemical potential,
\begin{equation}
\tilde P = \rho(\tilde\mu-\frac{F}{A})\,. 
\label{eq:pressmicro}
\end{equation}

The equalities $\tilde\mu=\mu$ and $\tilde P=P$ should hold in theory, but in practice numerical calculations always bring in errors. We have checked that our results are for the most part thermodynamically consistent. We have found discrepancies above saturation density which depend on the strength of the three-body force. 
These discrepancies are below $\sim2\%$ in relative terms at saturation density and $T=15$ MeV, and affect mostly results above saturation density. Improvements over the evaluation of $\langle\hat W\rangle$, i.e. including terms beyond Hartree-Fock, could help reduce these errors but are beyond the scope of this work.

In the following, unless otherwise stated, we rely on chemical potentials and pressures obtained from the derivatives of the free-energy as given in Eqs.~(\ref{eq:mumacro}) and (\ref{eq:pressmacro}).  With this procedure, these quantities automatically fulfill the Hugenholtz-Van Hove theorem~\cite{Hugen1958}, as given in Eq.~\eqref{eq:pressmicro} and the microscopic chemical potential, $\tilde \mu$, becomes an auxiliary quantity that is not needed for the solution of the liquid-gas transition.
We provide details on how the derivatives are computed in Sec.~\ref{subsec:therm_quant}. 
We also note that the virial expansion at high temperatures and low densities is an approximation of the  SCGF ladder approach \cite{Schmidt1990,Horowitz2006}. Our results therefore reproduce  the virial expansion in this regime \cite{Rios2009}.

\subsection{Finite temperature Brueckner-Hartree-Fock}
\label{finiteT_BHF}

We will compare the full ladder calculations of SCGFs to the G-matrix calculation within the BHF approach~\cite{Rios2005}. We expect that this will allow us to estimate uncertainties related to the many-body approximation. 
The Bloch-de Dominicis formalism is the generalization of the BHF approach to finite temperature \cite{bloch58a,*bloch58b,*bloch59a,*bloch59b}. We work within the low-temperature approximation of the Bloch-de Dominicis formalism, which is a direct extension of the zero-temperature BHF approach where all momentum step functions are replaced by  finite-temperature Fermi-Dirac occupation numbers. We refer the reader to Refs.~\cite{Baldo1999} for further details. This approach has been employed in most traditional finite-temperature BHF calculations \cite{bombaci94,zuo03,zuo06}. 

The method can also be obtained as a simplification of the SCGF approach in the quasi-particle limit, when only particle-particle intermediate states are considered \cite{Rios2005}. 
In other words, the BHF approximation does not include hole-hole rescattering in the in-medium interaction and  considers only a delta-like spectral function. 
We note that these two approaches are non-perturbative, include diagram resummation to all orders and have well-defined zero-temperature counterparts \cite{Wellenhofer2018}.
In order to establish as close a comparison as possible, we use partial-waves matrix elements of the 2N and 3N up to $J=8$ both in the Hartree-Fock and in the energy-dependent part of the self-energies for both approaches. 

\begin{table*}
\caption{
Summary of the parameters in the different chiral 2N  and 3N forces used in this work. 
These include: the cutoff of the SRG evolution on the 2N force, $\lambda_{\rm SRG}$ (column 2, where applicable); the cutoff $\Lambda_{3N}$ (column 3) and exponent $n$ (column 4) in the regulator function of the 3N force (see text); and the associated low-energy constants, $c_i$ (from column 5 to 9). Parameters are taken from Refs.~\cite{Hebeler2011,Klos2018,Ekstroem2015}.}
\begin{ruledtabular}
\begin{tabular}{p{2.25cm} p{1.5cm} p{1.2cm} p{0.5cm} p{2cm} p{2cm} p{2cm} p{2cm} p{2cm}}
 & $\lambda_{\rm SRG}\,  [{\rm fm}^{-1}]$ & $\Lambda_{3N}$ & $n$ & $c_1 \, [{\rm GeV}^{-1}]$ & $c_3\, [{\rm GeV}^{-1}]$ & $c_4\, [{\rm GeV}^{-1}]$ & $c_D$ & $c_E$ \\ \hline
 N3LO$_{\rm SRG18}$ \cite{Hebeler2011} & 1.8 & 2.0\,fm$^{-1}$ & 4 & -0.81 & -3.2 & 5.4 & 1.264 & -0.120 \\
 N3LO$_{\rm SRG20}$ \cite{Hebeler2011}  & 2.0 & 2.0\,fm$^{-1}$ & 4 & -0.81 & -3.2 & 5.4 & 1.271 & -0.131 \\
 N3LO$_{\rm SRG28}$  \cite{Hebeler2011} & 2.8 & 2.0\,fm$^{-1}$ &4 & -0.81 & -3.2 & 5.4 & 1.278 & -0.078 \\
 N2LO$_{\rm sat}$ \cite{Ekstroem2015} & / & 450\,MeV & 3 & -1.12152120 & -3.92500586 & 3.76568716 & 0.81680589 & -0.03957471 \\
N3LO$_{\rm full}$ \cite{Klos2018} & / & 500\,MeV & 3 & -0.81 & -3.2 & 5.4 & 0.339 & -0.610 \\ 
\end{tabular}
\end{ruledtabular}
\label{table:chiral_hamiltonian}
\end{table*}

The BHF method is not thermodynamically consistent~\cite{Rios2008}, so to solve the liquid-gas phase transition one necessarily needs to derive the macroscopic thermodynamic quantities. 
Previous comparisons between BHF and SCGF indicate that the role of many-body correlations is important for the phase transition. Generally speaking, for a given density and temperature, BHF calculations are more attractive than SCGF results. In turn, this means that a higher temperature is necessary to achieve the critical point \cite{Rios2008}. Here, we reassess these results with a larger set of Hamiltonians that, importantly, include 3N forces. 
We note that 3N forces are included in the BHF approach following the same scheme as in the the SCGF method, discussed in Sec.~\ref{subsec:scgf_form}. Previous non-perturbative BHF and SCGF calculations of the phase transition have implemented other averaging procedures for the 3N force \cite{zuo03,Soma2009}. In these past studies, a substantial influence of 3N forces on the critical properties of the liquid-gas phase transition has been found. In particular, the repulsive effect of 3N in the equation of state brings in a substantial reduction of the critical temperature.

\subsection{Chiral Hamiltonians}
\label{subsec:chiral_ham}
\iffalse
\begin{itemize}
\item Chiral EFT
\item 5 Hamiltonians in use
\item details on 3NFs
\end{itemize}
\fi

In the chiral EFT realization of low-energy QCD, the effective degrees of freedom that come into play are nucleons and pions. 
One could additionally consider the $\Delta$-isobar degree of freedom \cite{Logoteta2016}, but in the present formulation we do not include it. The high-energy physics which is integrated out of the theory is encoded in contact terms. The strength of these terms is given by specific low-energy constants (LECs) which are traditionally fit to scattering phase-shifts or few-body properties. In the present calculations, we make use of five different parametrizations of chiral Hamiltonians which cover a range of LECs values and orders in the chiral expansion for the 2N force. All our 3N forces are constructed at N2LO. Overall, our analysis should provide estimates of the systematic theoretical uncertainty associated with different  constructions of the 2N force and of the fitting of the LECs for both the 2N and the 3N contributions. 

Table~\ref{table:chiral_hamiltonian} gives all the details of the five Hamiltonians we are using for the present calculations. For the 2N part, three Hamiltonians are obtained from different similarity renormalization group (SRG) transformations of the Entem-Machleidt 2N Hamiltonian (EM500) up to next-to-next-to-next-to leading order (N3LO)~\cite{Entem2003}. The three different cases are obtained with different cutoffs of the SRG evolution, $\lambda_{\rm SRG}$. We note that, in keeping with previous studies, we ignore the effect of SRG-induced 3N forces~\cite{Hebeler2011}.  We also consider the case of two unevolved 2N forces: 
 a consistent (2N+3N)-force up to N2LO constructed using the practical optimization using no derivatives (for squares) algorithm (the so called N2LO$_{\rm sat}$ potential~\cite{Ekstroem2015}) and the N3LO EM500, which we dub N3LO$_\text{full}$. 

We complement each of these 2N interactions with a 3N force at N2LO~\cite{VanKolck1994,Epelbaum2002,Carbone2014}. 
The 3N force is included in the many-body calculation via the correlated average discussed in detail in Sec.~\ref{subsec:scgf_form} and Ref.~\cite{Carbone2014}\footnote{Because of the finite-temperature correlated average, we have corrected the scalar term multiplied by $c_3$ in Eq.~(A11) of Ref.~\cite{Carbone2014}, e.g. $8k_{\rm F}^3\rightarrow 8\rho6\pi^2/\nu$.}. This 3N force is regularized with a non-local regulator $f({\bf p},{\bf q})=\exp[-(p^2+3q^2/4)/\Lambda_{3N}^2]^n$, where ${\bf p}$ and ${\bf q}$ are the corresponding Jacobi momenta. 
The LECs $c_D$ and $c_E$ of the 3N force are associated to the one-pion exchange and contact 3N terms respectively and need to be fit to few-body properties. These constants are obtained using different strategies for each 2N force. For the SRG evolved Hamiltonians, fits are performed to reproduce the $^3$H binding energy and the $^4$He matter radius~\cite{Hebeler2011}. For N3LO$_\text{full}$, the coupling constants are fit to reproduce the triton $\beta$-decay and the $^3$H binding energy~\cite{Klos2018}. For the N2LO$_{\rm sat}$ case, the 2N and 3N Hamiltonians are simultaneously optimized, including binding energies and radii of $^3$H and $^{3,4}$He, as well as properties of carbon and oxygen isotopes~\cite{Ekstroem2015}. These determinations span a large space of possibilities in the fitting criteria and therefore provide a good starting point for uncertainty quantification. 
We give in Table~\ref{table:chiral_hamiltonian} the values of pion-nucleon $c_i$'s; the values of the $c_D$ and $c_E$ LECs, and also details of cutoffs and exponents in the regularization of each chiral interaction.

\section{Results}
\label{sec:results}

\begin{figure*}[t!]
\includegraphics[scale=0.6]{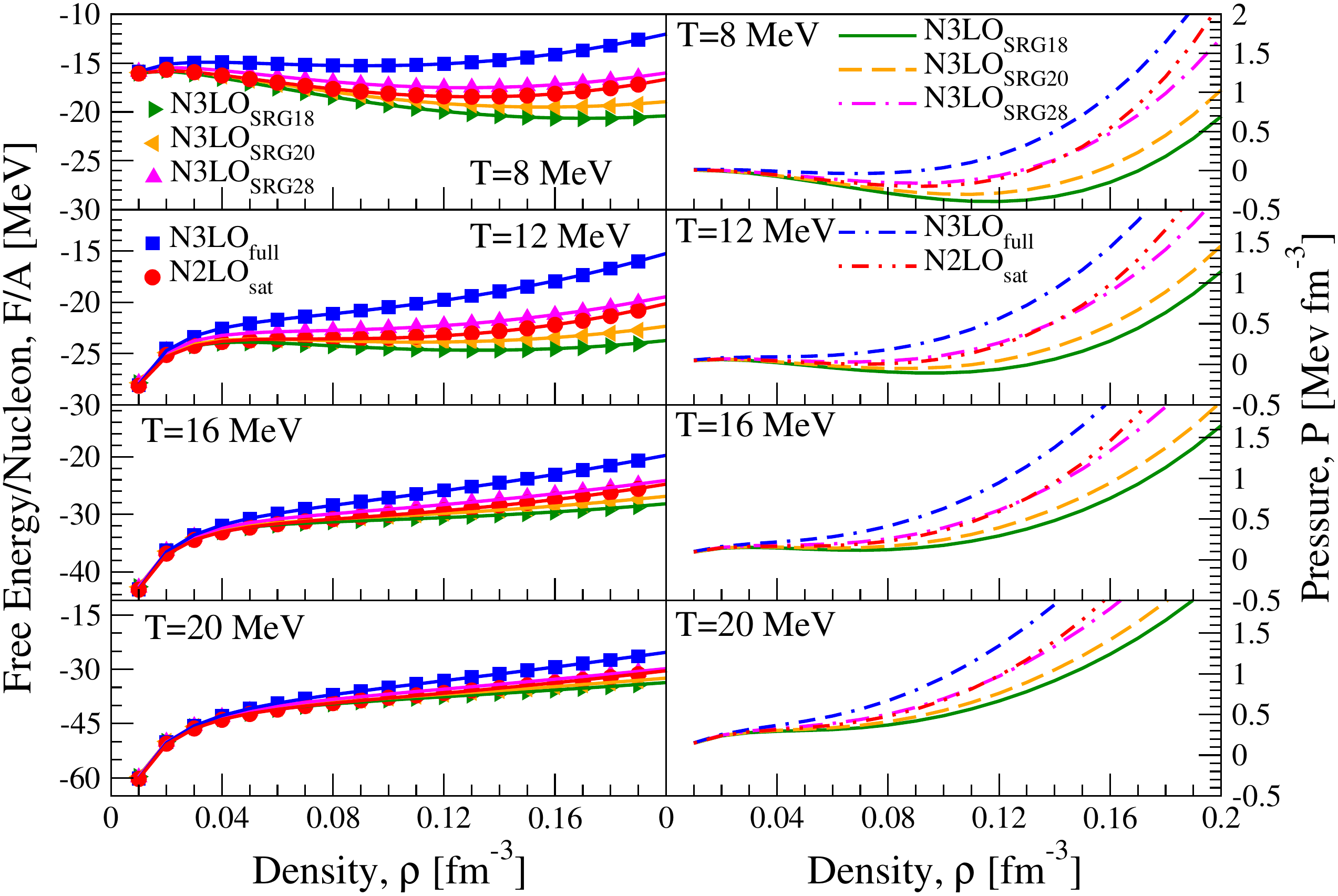}
\caption{Left panels: free energy as a function of density for temperatures $T=8, 12, 16$ and $20$ MeV (from top to bottom) employing the five Hamiltonians of Table~\ref{table:chiral_hamiltonian} within the SCGF method. Symbols represent SCGF calculated data, while lines are the fits using Eq.~\eqref{eq:freefit}. Right panels: pressure in the same conditions. The pressure is obtained as a derivative of the free-energy based on Eq.~\eqref{eq:pressmacro}.}
\label{fig:free_press_temp}
\end{figure*}

\subsection{Thermodynamical properties of nuclear matter}
\label{subsec:therm_quant}

We start our discussion by providing a set of results for the free energy, chemical potential and pressure of symmetric nuclear matter as a function of density and temperature in the regime of interest for the liquid-gas phase transition. We use the five different Hamiltonians described in Sec.~\ref{subsec:chiral_ham}. To perform the derivatives of Eqs.~(\ref{eq:mumacro}) and (\ref{eq:pressmacro}), we fit our calculated data for the free-energy at a given temperature with a function similar to that proposed in Ref.~\cite{Drischler2016}. We also include a constant term which depends on temperature,
\begin{equation}
\frac{F}{A}(\rho,T)=a(T)+\sum_{\nu=2,3,4,5,6}a_\nu(T)\left(\frac{\rho}{\rho_{\rm sat}}\right)^{\nu/3}\,,
\label{eq:freefit}
\end{equation}
and we use $\rho_{\rm sat}=0.16$ fm$^{-3}$. We have tested that different fits of the free energy lead to numerical errors which are small and well within the final many-body calculation uncertainty band. 

\begin{figure*}[t!]
\includegraphics[scale=0.6]{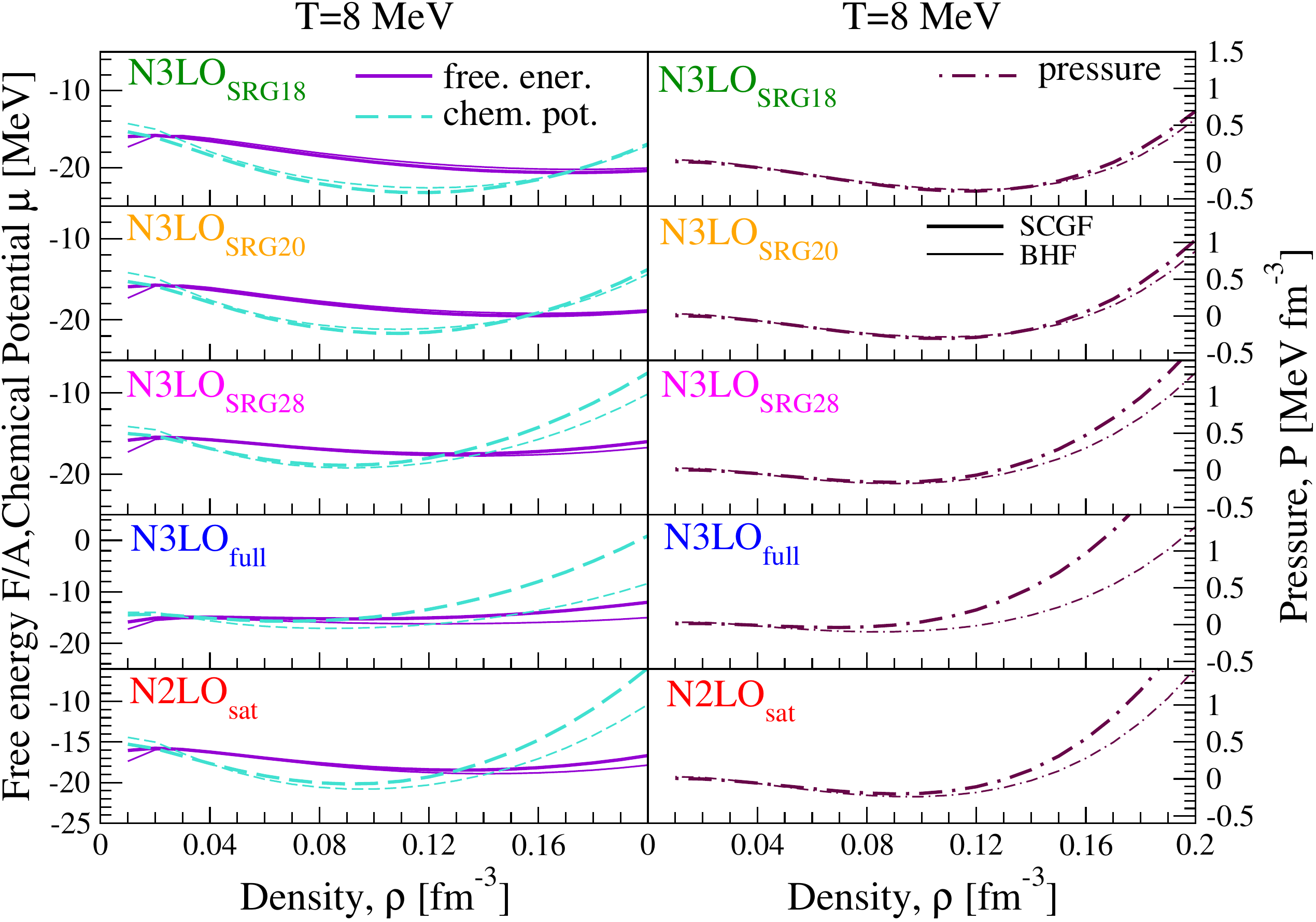}
\caption{Left panels: free energy (solid lines) and chemical potential (dashed lines) as a function of density at $T=8$ MeV. Thick lines correspond to SCGF results, while thinner lines show BHF data (see text for details). The panels from top to bottom display results for the five different Hamiltonians of Table~\ref{table:chiral_hamiltonian}. Right panels: pressure in the same conditions.}
\label{fig:mb_dep}
\end{figure*}

We calculate the free energy per nucleon and pressure as functions of baryonic density employing the five Hamiltonians presented in Table~\ref{table:chiral_hamiltonian}, for several temperatures. We show in the  panels of Fig.~\ref{fig:free_press_temp} the results obtained within the SCGF method for four specific temperatures, $T=8,12,16$ and $20$ MeV (from top to bottom respectively). 
For the free energy (left panels), we provide two different sets of results. Points represent the SCGF calculations of the free energy per particle. The corresponding solid lines represent the fits of Eq.~\eqref{eq:freefit} to the same set of data. We find a very good agreement between the two. 

For the lowest temperatures (Fig.~\ref{fig:free_press_temp}, top two left panels), the free energy presents a non-convex curvature region which corresponds to a mechanically unstable phase for matter, which signals the appearance of a first-order phase transition. This transition is reflected also in the pressure as a function of density (right panels). At low temperatures, there are regions where two different values of density provide the same value for the pressure. This defines the coexistence in equilibrium of two different phases of nuclear matter, a gas (at low densities) and a liquid (at high densities). As the temperature increases, one finds that coexistence gradually disappears. In fact, at $T=20$ MeV (bottom panels), all predictions provide a one-to-one pressure-density relation. This shows that $T=20$ MeV is above the critical temperature, $T_c$. 

The results associated to the five different Hamiltonians give an estimate of the theoretical uncertainty related to the construction of the chiral nuclear forces. We first note that the predictions from different Hamiltonians are ordered in well-defined trends that fall within relatively narrow bands. For instance, all chiral forces predict free energies which are within about $5$ MeV of each other across the whole density range, except for N3LO$_{\rm full}$. The latter is substantially more repulsive at all densities and temperatures. We also find that, as the temperature increases, the relative differences remain constant. 
Generally speaking, more evolved forces provide more attractive results. N3LO$_{\rm SRG18}$ and N3LO$_{\rm SRG20}$ are the most attractive ones, with a relative deviation in free-energies of $\sim 1$ MeV at $\rho_{\rm sat}$. N3LO$_{\rm SRG28}$ free-energy results are a bit more repulsive, $\sim 1$ MeV at  $\rho_{\rm sat}$, than N2LO$_{\rm sat}$ predictions. The latter is also about an MeV more repulsive than the N3LO$_{\rm SRG20}$ results. There is however a relatively large gap between N3LO$_{\rm SRG28}$ and the N3LO$_{\rm full}$ results, of the order of $\sim 3$ MeV. To some extent this is unexpected, since the SRG cutoff is relatively large, $\lambda_{\rm SRG}=2.8$ fm$^{-1}$, and one would have expected small differences between the two. We note, however, that high-momentum correlations are well described in our approach and our results should quantitatively account for their presence. We note that the relative differences in free energies can increase with density, due to the different strength of 3N forces for each chiral Hamiltonian. However, the relative differences remain constant with increasing temperature. This indicates that the trends imposed by the Hamiltoninan are rather independent of temperature.

The behavior that we have just described for the free energy has clear parallels in the pressure, shown in the right panels of Fig.~\ref{fig:free_press_temp}. We find that the more evolved Hamiltonians predict generally lower pressures across the whole density and temperature regime. There is a clear separation between N3LO$_{\rm SRG18}$ and N3LO$_{\rm SRG20}$ as in the free-energy case, whereas N3LO$_{\rm SRG28}$ and N2LO$_{\rm sat}$ are close to each other. In fact, the density dependence is such that the predictions for these two forces cross at a given point at all temperatures. 
The repulsion of N3LO$_{\rm full}$ translates into a distinctly larger pressure compared to all other forces. 

We note that qualitatively similar results would be obtained in the BHF approximation. To discuss the dependence of these results on the many-body method, we focus on a single temperature and provide results as a function of density.
The left column of Fig.~\ref{fig:mb_dep} shows results for chemical potential and free energy at $T=8$ MeV. The pressure is displayed in the right panels. Each panel within the left and right columns corresponds to one of the five Hamiltonians described in Table~\ref{table:chiral_hamiltonian}.

The uncertainty associated to the many-body approximation can be estimated by comparing the SCGF results (thick lines) to the BHF calculations (thin lines). 
We identify three clear trends. First, the many-body method dependence of the results is stronger as the density increases. 
This is expected, because the effect of hole-hole scattering states, considered in SCGF but not in BHF, is stronger as the available phase space becomes larger with density \cite{riosphd}. Second, the inclusion of hole-hole scattering induces mainly a repulsive effect. Where the effect is significant at large densities, SCGF predictions for the free energy and the chemical potential are significantly less attractive than their BHF counterparts. In turn, this results in an overall larger pressure for SCGF predictions. Third, the size of the many-body uncertainty depends on the chiral force under consideration. For the two most evolved forces  N3LO$_{\rm SRG18}$ and N3LO$_{\rm SRG20}$, the differences between the approximations are within half an MeV for both $F/A$ and $\mu$ across the whole density range. For the evolved N3LO$_{\rm SRG18}$, the differences are similarly small but rise as density increases, reaching a maximum of $\sim 2.5$ MeV for the chemical potential at the highest density shown, $\rho=0.20$ fm$^{-3}$. In contrast, even at lower densities, there is a significant difference between the SCGF and BHF results for both N2LO$_{\rm sat}$ and, especially, N3LO$_{\rm full}$. Evolved potentials are soft, and one expects them to be perturbative and converge at a relatively low order in the many-body expansion \cite{Drischler2016}. In that case, the differences between two non-perturbative approaches that treat low orders similarly should be small, just as we find here. Having said that, we note that the many-body results are not entirely independent of the SRG scale, which is an indication that higher-order induced many-body forces may play a role in infinite matter. By providing predictions for several Hamiltonians, we expect to account for the missing physics of these induced forces. 

The uncertainty associated to the many-body truncation in Fig.~\ref{fig:mb_dep} is smaller than the one arising from the use of different chiral Hamiltonians. For instance, at saturation density $\rho_\text{sat}=0.16$ fm$^{-3}$, the maximum difference between SCGF and BHF free energies at $T=8$ MeV is encountered for the N3LO$_{\rm full}$ interaction and amounts to $\sim1.8$ MeV. In contrast, at this same temperature and density, the different Hamiltonians provide results within a window of $\sim 6.5$ MeV. The phase transition explores the subsaturation region so uncertainties there are expected to be smaller in absolute scale. However, the Hamiltonian uncertainties are likely to dominate in relative terms the error budget too. 
This is a consequence of the uncertainty in constraining reliably the LECs which characterize the three-body chiral contributions~\cite{Klos2018}.

\subsection{Liquid-gas phase transition for N2LO$_{\rm sat}$}
\label{subsec:nnlo_sat_lg}
\label{subsec:liquid-gas}

\begin{figure}
   \centering
    
    \begin{subfigure}[t]{0.45\textwidth}
       \includegraphics[width=\textwidth]{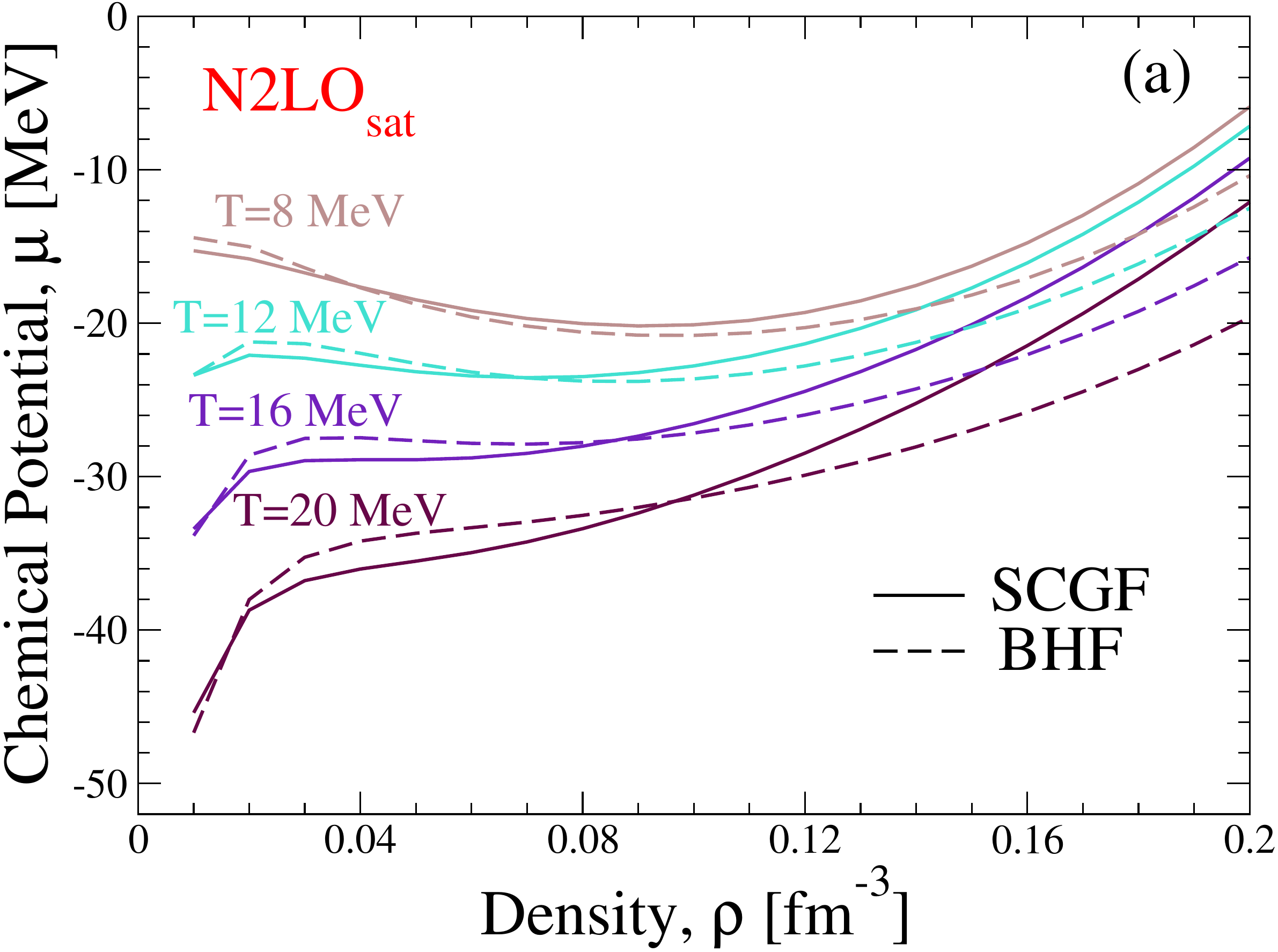}
       \label{subfig:chempot_nnlosat}
     \end{subfigure}

    \begin{subfigure}[t]{0.45\textwidth}
        \includegraphics[width=\textwidth]{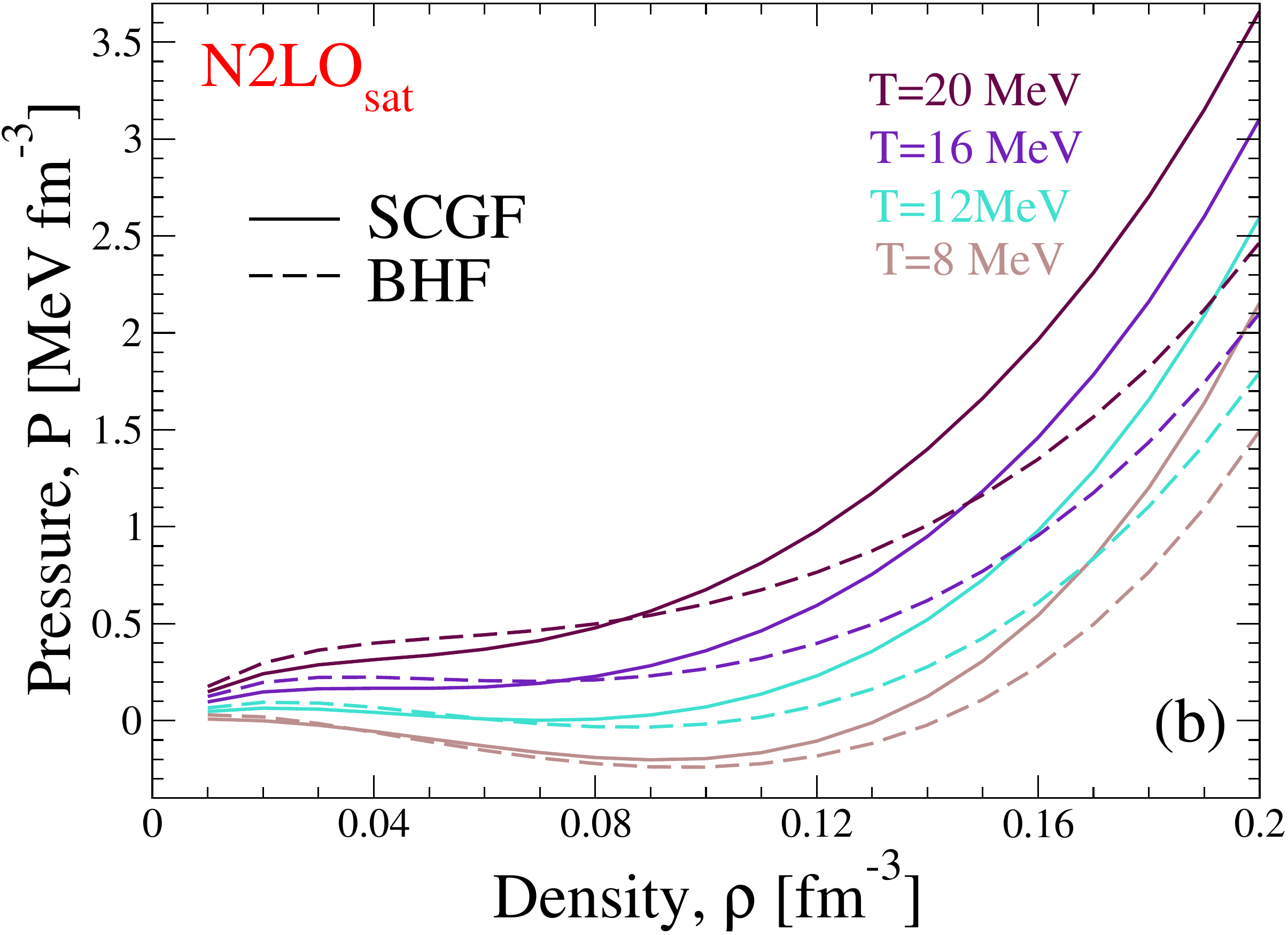}
        \label{subfig:press_nnlosat}
    \end{subfigure}

    \begin{subfigure}[t]{0.45\textwidth}
        \includegraphics[width=\textwidth]{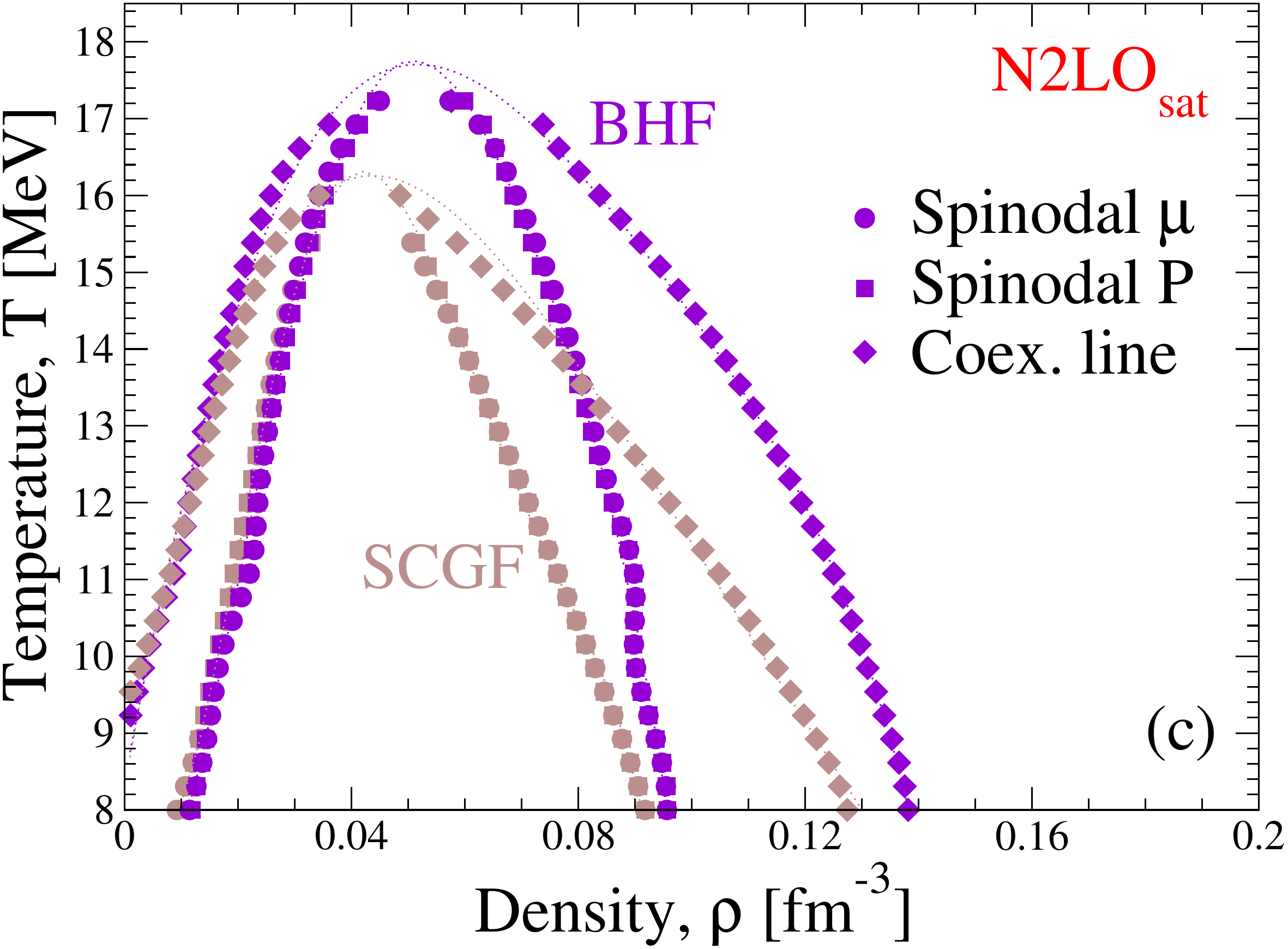}
        \label{subfig:liquid-gas_nnlosat}
    \end{subfigure}
       
    \caption{
      (a) Chemical potential and (b) pressure as a function of density for the N2LO$_{\rm sat}$ potential (2N+3N) at four different temperatures (see legend). Solid lines correspond to SCGF results and dashed lines refer to BHF calculations. (c) Coexistence (diamonds) and spinodal (circles and squares) lines as a function of density for the same force. Spinodal lines for both chemical potential (circles) and pressure (squares) are plotted.}
    \label{fig:nnlo_sat}
\end{figure}

We now analyze the liquid-gas phase transition for the specific case of the N2LO$_{\rm sat}$ potential. A qualitatively similar discussion would apply to the other Hamiltonians of Table~\ref{table:chiral_hamiltonian}. 
In Fig.~\ref{fig:nnlo_sat}(a) and ~\ref{fig:nnlo_sat}(b) we show the chemical potential and pressure as functions of density for a selection of temperatures in the range $T=8$ to $20$ MeV for both SCGF (solid lines) and BHF (dashed lines). In Fig.~\ref{fig:nnlo_sat}(a) we see that, at low densities, the chemical potential becomes more and more attractive as the temperature increases and the density decreases. This is in agreement with the logarithmic density dependence of this quantity in the classical limit. As the density increases, however, the system becomes more degenerate and temperature effects become less relevant. As a consequence, the chemical potentials at different temperatures tend to have closer values. In addition, the chemical potential presents a local minimum at low temperatures. This minimum disappears above the critical temperature, $T_c$, which this data suggest is around $16$ MeV.

In a similar fashion, Fig.~\ref{fig:nnlo_sat}(b) illustrates that at low temperatures the pressure has regions of negative slopes, associated to a mechanical instability, up to a given local minimum. This instability region shrinks as temperature increases and eventually disappears when $T_c$ is reached. The presence of these minima in both the chemical potential and the pressure is the reason why one can find two points with different density but the same value of $\mu$ and $P$. The critical temperature $T_c$ signals the point in temperature where the coexistence of the liquid and gas phases is no longer possible. Above this temperature, the chemical potential and the pressure are always a monotonically increasing function of density.

The uncertainty in the many-body truncation follows the trends discussed in Sec.~\ref{subsec:therm_quant}. BHF results are more attractive than their SCGF counterparts. The differences between the two approaches depend on density (as already mentioned) and temperature. For instance, at $\rho=0.16$ fm$^{-3}$, the SCGF pressure at $T=8$ MeV differs from the BHF one by $0.3$ MeV fm$^{-3}$. At $T=20$ MeV, this difference becomes $0.6$ MeV fm$^{-3}$, e.g. the discrepancy has doubled. The differences rise with density, and we find at $\rho=0.20$ fm$^{-3}$ a discrepancy of $0.7$ MeV fm$^{-3}$ for $T=8$ MeV, which becomes $1.2$ MeV fm$^{-3}$ at $T=20$ MeV. We note that these trends are different from those associated to the different Hamiltonians which, in relative terms, stay qualitatively constant as a function of temperature. 

The phase diagram of the liquid-gas phase transition is described in Fig.~\ref{fig:nnlo_sat}(c). Results are provided for both SCGF and BHF predictions. 
The coexistence line (diamonds) defines at each temperature the two points in density which fulfill the condition $\mu(\rho_g)=\mu(\rho_l)$ and $P(\rho_g)=P(\rho_l)$. Here, $\rho_g$ is the density of the gas phase and $\rho_l$ is the density of the liquid phase. In between these points, the system coexists in a liquid and gas phase.  
To solve the coexistence equations, we calculate the chemical potential and pressure as derivatives of the free-energy at a fixed set of densities and temperatures spaced by $0.01$ fm$^{-3}$ and $1$ MeV, respectively.
We then perform two-dimensional density and temperature fits using polynomial interpolations to access a denser grid in both quantities. The polynomials can also be used to solve the conditions for the phases coexistence, i.e. $\mu(\rho_g)=\mu(\rho_l)$ and $P(\rho_g)=P(\rho_l)$. Numerical errors associated with these interpolations are negligible. 

The spinodal line corresponds to an area of mechanical or chemical instability. This is in fact the region in the phase diagram where the derivatives of both the chemical potential and the pressure become  negative, thus violating the thermodynamical stability criteria. To define the spinodal lines, one then searches for the maxima and minima of both $\mu$ and $P$, $\frac{\partial\mu}{\partial\rho_{g}}=\frac{\partial\mu}{\partial\rho_{l}}=0$ and $\frac{\partial P}{\partial\rho_g}=\frac{\partial P}{\partial\rho_l}=0$. We can thus obtain two spinodal lines: one representing chemical and the other one mechanical instabilities. These two lines are shown in Fig.~\ref{fig:nnlo_sat}(c) with circles (chemical potential) and squares (pressure), respectively. 
These two lines should coincide with one another, and in fact they do as seen in  Fig.~\ref{fig:nnlo_sat}(c). We take this agreement as a further confirmation of the numerical procedure that we have set up to study the phase transition. 

The critical temperature $T_c$ can be estimated in three different ways: as the maximum of the chemical potential spinodal line; of the pressure spinodal line; or of the corresponding coexistence line. The small dotted lines that join the points in Fig.~\ref{fig:nnlo_sat}(c) represent polynomial interpolations of all these data. We use the maxima of these polynomials to provide an estimate of the critical point ($T_c$~,~$\rho_{\rm c}$), which should be the same when extracted from either of the three estimates. We find that indeed these three predictions agree well with each other. As seen in Fig.~\ref{fig:nnlo_sat}(c), the three approaches yield almost undistinguishable maxima for both the BHF and the SCGF results. A similar discussion applies for the other four Hamiltonians. The critical point thus obtained is reported in Table~\ref{table:values}.

Furthermore, Fig.~\ref{fig:nnlo_sat}(c) illustrates the general behavior of the many-body truncation dependence of our results. The phase diagrams for both the BHF and SCGF are qualitatively similar, but presents also some quantitative differences. The critical point occurs at a slightly higher temperature for BHF compared to SCGF, as expected from the overall repulsive nature of hole-hole correlations. Well below the critical temperature, the coexistence and spinodal lines on the low-density gas side are very similar for both SCGF and BHF. These only deviate from each other about $2$ MeV below $T_c$. This low-density region is relatively insensitive to many-body differences as expected on the grounds of the virial expansion \cite{Horowitz2006}, which both BHF and SCGF calculations incorporate by construction. 

The liquid boundary of the phase diagram occurs at higher densities and therefore we expect it to be more sensitive to many-body correlations. We do find that the liquid coexistence line occurs at larger densities for BHF compared to SCGF. The same holds for the spinodal lines at high densities. Compared to previous results obtained with 2N-forces only \cite{Rios2008}, the critical temperature is relatively lower for both BHF and SCGF. However, the differences between the two many-body approaches are relatively similar here and in Ref.~\cite{Rios2008}, indicating that 3N forces do not qualitatively change the picture when it comes to many-body uncertainties. 

\begin{figure*}
   \centering
    
    \begin{subfigure}[t]{0.45\textwidth}
       \includegraphics[width=\textwidth]{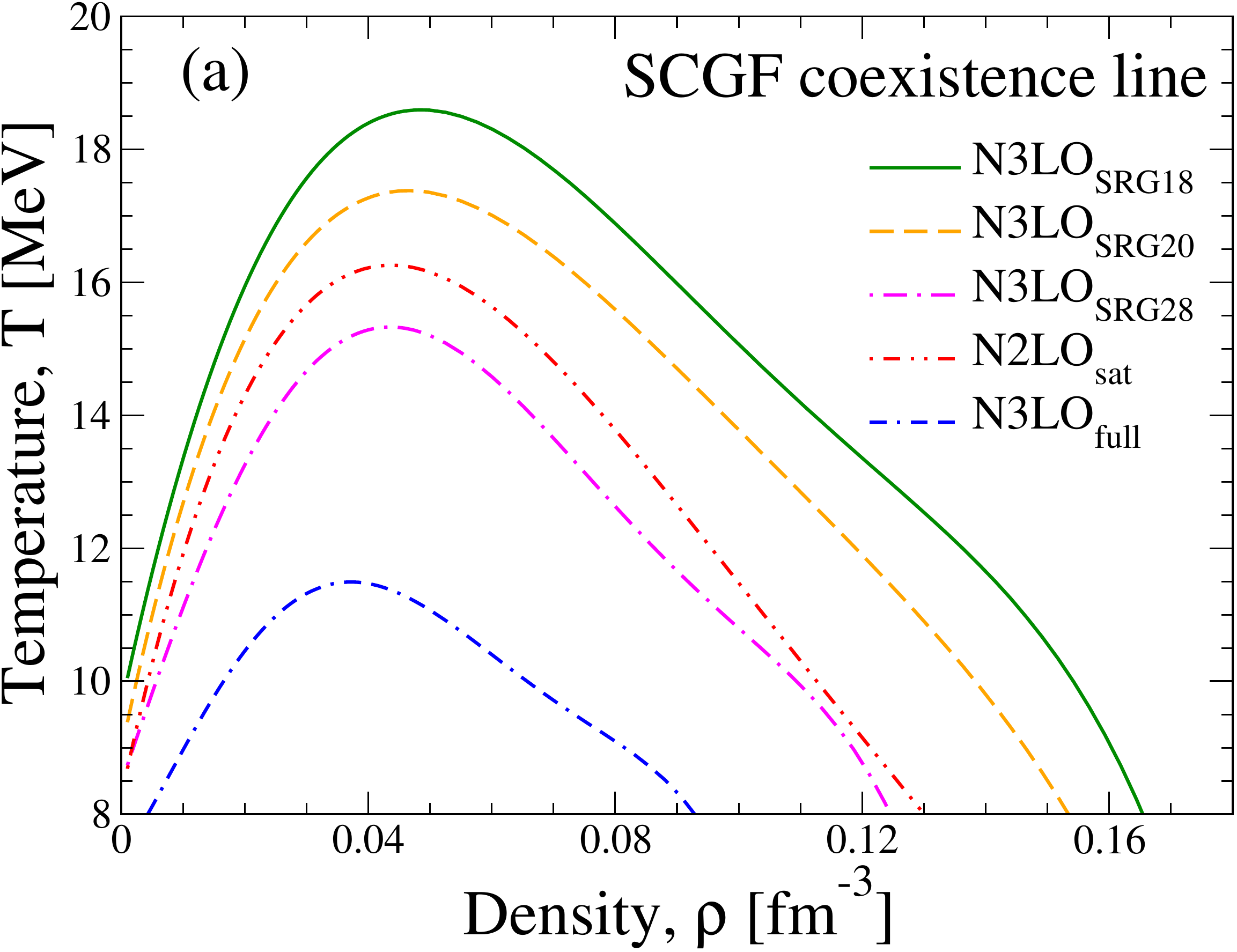}
        \label{subfig:coex_scgf}
     \end{subfigure}~~~~
     \begin{subfigure}[t]{0.45\textwidth}
        \includegraphics[width=\textwidth]{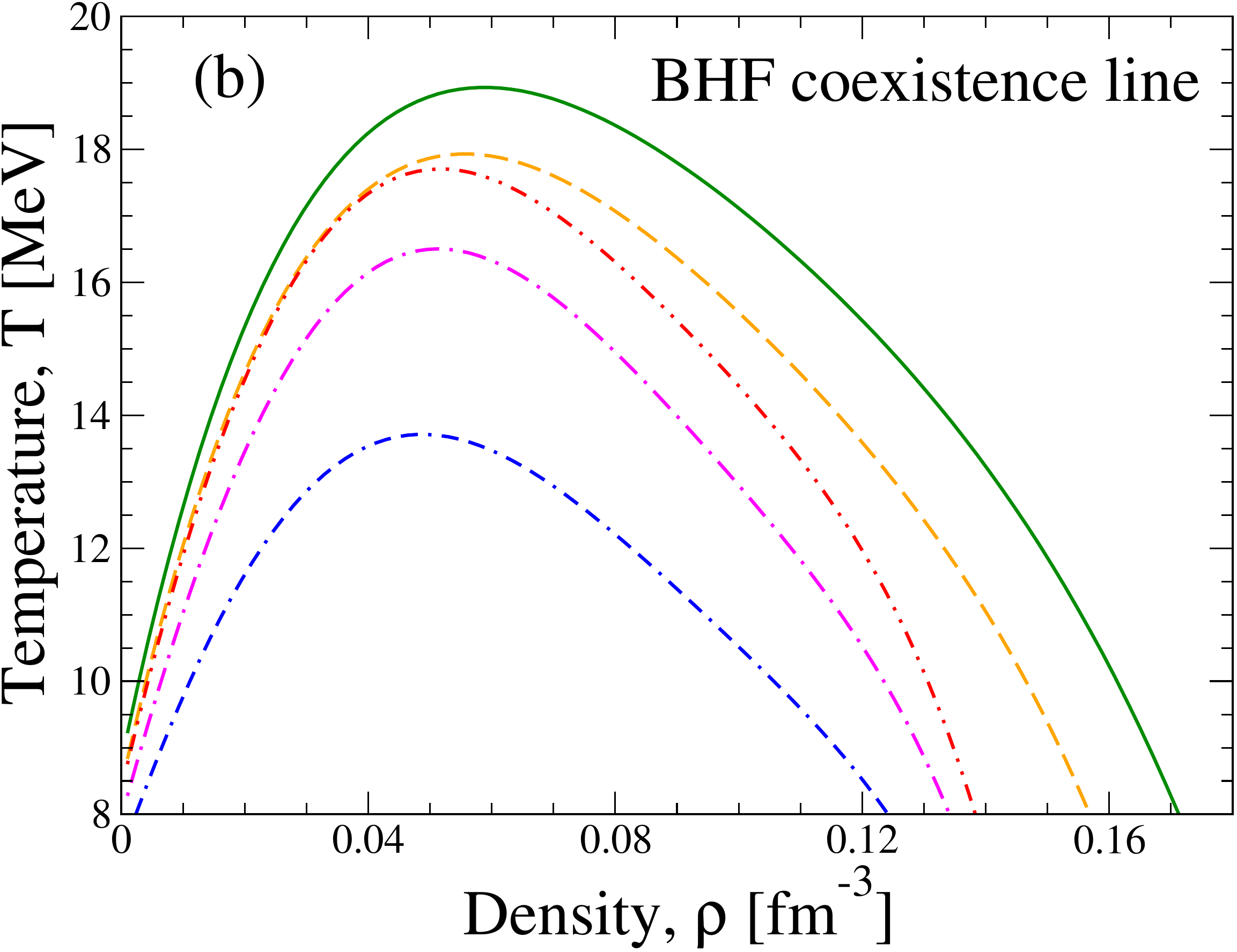}
        \label{subfig:coex_bhf}
    \end{subfigure}
       
    \caption{Coexistence lines using the five Hamiltonians given in Table~\ref{table:chiral_hamiltonian} within (a) the SCGF and (b) the BHF methods.}
    \label{fig:coex_line}
\end{figure*}

\subsection{Coexistence line and critical temperature}
\label{subsec:coex_line}
\iffalse
\begin{itemize}
\item the coexistence line for all 5 Hamiltonians
\item relation $Tc/\rho_0$
\end{itemize}
\fi

We now extend the analysis of the many-body truncation dependence of our results to the five chiral forces of Table~\ref{table:chiral_hamiltonian}. Figure~\ref{fig:coex_line} shows the coexistence lines obtained for all potentials: for Fig.~\ref{fig:coex_line}(a), the SCGF approximation, and Fig.~\ref{fig:coex_line}(b), the BHF approximation. 
We have applied the same procedure discussed in the previous section to solve for the phase coexistence. 
The Hamiltonians which give a more attractive free energy and a lower pressure, such as N3LO$_{\rm SRG18}$ (solid lines), lead to coexistence lines which cover a wider area in the phase diagram. In contrast, the more repulsive forces, such as N3LO$_{\rm full}$ (double-dash-dotted line), cover a substantially smaller region of the density-temperature plane. There is a clear separation in the phase diagram of all forces, which give a coexistence line (and thus a phase diagram) which lies in between the limits established by N3LO$_{\rm SRG18}$ at the top and N3LO$_{\rm full}$ at the bottom. 

The critical density and temperature for each Hamiltonian in the two many-body approximations are reported in columns 2 and 3 of Table~\ref{table:values}. 
The Hamiltonian dependence of the critical point is a direct reflection of the overall attractive or repulsive nature of the free-energy. We note that the ordering in terms of free energy described in the left panels of Fig.~\ref{fig:free_press_temp} is in one-to-one correspondence to the phase diagrams presented in  Figs.~\ref{fig:coex_line} and the critical properties of Table~\ref{table:values}. 

There is a substantial dependence of the critical temperature on Hamiltonians, spanning over $7$ ($5$) MeV in the SCGF (BHF) case. A previous study using 2N phase-shift equivalent forces Av18 and CD-Bonn found a similar difference in critical temperatures between the two potentials and for the two many-body methods \cite{Rios2008}. If N3LO$_{\rm full}$ predictions are removed, however, our critical temperature results reduce to a more conservative span of $T_c=16.3-18.9$ MeV. In fact, computing a mean and a standard deviation from all the results of Table~\ref{table:values}, including N3LO$_{\rm full}$, we estimate the critical temperature to be  in the range $T_c=16.4 \pm 2.3$ MeV. 

The trends and figures indicate that N3LO$_{\rm full}$ is a relative outlier with a very low critical temperature. Similar low critical temperatures of order $T_c=11-12$ MeV have been reported in the past in relativistic many-body calculations ~\cite{terhaar86,huber98} as well as the 2N+3N force SCGF calculations of Ref.~\cite{Soma2009}. In the latter case, 3N forces were included without explicitly differentiating one- and two-body effective interactions, which provided an overall very large pressure, leading in turn to a low critical temperature. If  N3LO$_{\rm full}$ is considered an outlier and removed from the average, we find that the mean is slightly higher and the standard deviation narrower, $T_c=17.3 \pm 1.2$.
The perturbative calculations of Ref.~\cite{Wellenhofer2014} using similar chiral 2N and 3N forces provide critical temperatures in the region $T_c=17.4-19.1$ MeV, in agreement with these results. We note that the N3LO results in these perturbative calculation does not include effective masses and is computed with different $c_D$ and $c_E$ constants so, unsurprisingly, a much larger critical temperature is obtained. 

The uncertainty associated to the many-body truncation can also be extracted from the results presented in Figs.~\ref{fig:coex_line} and Table~\ref{table:values}. As discussed already in Sec.~\ref{subsec:nnlo_sat_lg}, all BHF predictions yield higher critical temperatures than their SCGF counterparts. This is because, as observed in Fig.~\ref{fig:mb_dep}, the chemical potentials are more attractive and the pressures are lower for the BHF calculation. In turn, these require larger temperatures to reach the critical point. As expected, the differences between many-body methods are smaller for softer, more perturbative forces. For N3LO$_{\rm SRG18}$, for instance, $T_c=18.9$ MeV for the BHF calculation whereas the SCGF results indicate $T_c=18.6$ MeV. In turn, less perturbative forces provide larger differences, and for N3LO$_{\rm full}$ the critical temperature goes from $T_c=13.7$ MeV in the BHF approach to $11.5$MeV in SCGF. 
These findings validate the results of Ref.~\cite{Rios2008}, where the decrease of $T_c$ due to hole-hole scattering was already discussed. 

We note that, if these results are representative of the many-body truncation uncertainty in the critical temperature, this uncertainty is, at most, $\approx 2$ MeV. In contrast, there is a spread of $5-7$ MeV in critical temperatures due to the Hamiltonians dependence, which clearly dominates the uncertainty budget. We note that this range in critical temperatures is similar to that obtained in theoretical analysis of relativistic and non-relativistic density functionals \cite{Rios2010,Lourenco2016}. Unlike those phenomenological approaches, however, our calculations are fully predictive and not based on fitting of the zero-temperature nuclear matter equation of state around saturation. 

While a detailed analysis of the effect of 3N forces on critical properties goes beyond the scope of our work, we note that 3N interactions are important in determining the critical properties. A BHF calculation of N3LO without 3N forces yields a critical temperature of the order $T_c \approx 27$ MeV, which is to be compared to the $13.7$ MeV obtained here. Previous calculations using phenomenological phase-shift equivalent potentials with 2N and with 2N+3N forces have also validated the importance of 3N forces. Reference~\cite{zuo03} reports a $3$ MeV decrease in $T_c$ when 3N forces are incorporated in a BHF calculation. The SCGF calculations of Ref.~\cite{Soma2009} show larger changes, of the order of $6-9$ MeV. 
 
\begin{table}
\caption{\label{table:values}Critical density, $\rho_c$, and temperature, $T_c$; saturation density, $\rho_0$, and energy, $E_0/A$; and effective mass at saturation density, $m_0^*/m$, for each of the five Hamiltonians considered in Table~\ref{table:chiral_hamiltonian}. We provide results for both the SCGF and BHF methods.}
\begin{ruledtabular}
\begin{tabular}{c c c c c c}
 SCGF & $\rho_c$ [fm$^{-3}$] & T$_c$ [MeV] & $\rho_0$ [fm$^{-3}$] & $\frac{E_0}{A}$ [MeV]& $\frac{m_0^*}{m}$  \\ \hline
 N3LO$_{\rm SRG18}$ & 0.048 & 18.6 & 0.19 & -17.6 & 0.83   \\
 N3LO$_{\rm SRG20}$ & 0.047 & 17.4 & 0.19 & -16.2 & 0.84 \\
 N3LO$_{\rm SRG28}$ & 0.043 & 15.3 & 0.16 & -13.7 & 0.88 \\
 N2LO$_{\rm sat}$ & 0.043 & 16.3 & 0.15 & -14.6 & 0.90  \\
 N3LO$_{\rm full}$ & 0.038 & 11.5 & 0.14 & -11.0 & 0.84 \\
 \hline\hline
BHF & $\rho_c$ [fm$^{-3}$] & T$_c$ [MeV] & $\rho_0$ [fm$^{-3}$] & $\frac{E_0}{A}$ [MeV]& $\frac{m_0^*}{m}$  \\ \hline
 N3LO$_{\rm SRG18}$ & 0.058 & 18.9 & 0.19 & -17.3 & 0.70  \\
 N3LO$_{\rm SRG20}$ & 0.056 & 17.9 & 0.18 & -16.0 & 0.73 \\
 N3LO$_{\rm SRG28}$ & 0.051 & 16.5 & 0.16 & -14.2 & 0.79 \\
 N2LO$_{\rm sat}$ & 0.051 & 17.7 & 0.16 & -15.2 & 0.82  \\
 N3LO$_{\rm full}$ & 0.048 & 13.7 & 0.16 & -12.6 & 0.76 \\
\end{tabular}
\end{ruledtabular}
\end{table}

\subsection{Connecting the critical and saturation points}
\label{subsec:crit_sat}

We have found a very clear trend connecting more attractive finite-temperature predictions of the free energy to larger critical temperatures. In a sense, this is natural: a larger thermal energy is needed to counteract the effect of a more attractive nuclear force before the system can dissolve at the critical point. 
In the zero-temperature limit, the free energy reduces to the energy per particle which, for nuclear matter, has a minimum around the saturation point. In fact, the zero-temperature liquid-gas latent heat is equal to the saturation energy \cite{Carbone2011}. According to the arguments above, one might expect that more attractive saturation points correlate with larger critical temperatures. Having access to five Hamiltonians and two different many-body methods, we now have enough data to test whether such correlation exists. 

\begin{figure}[t!]
\includegraphics[scale=0.36]{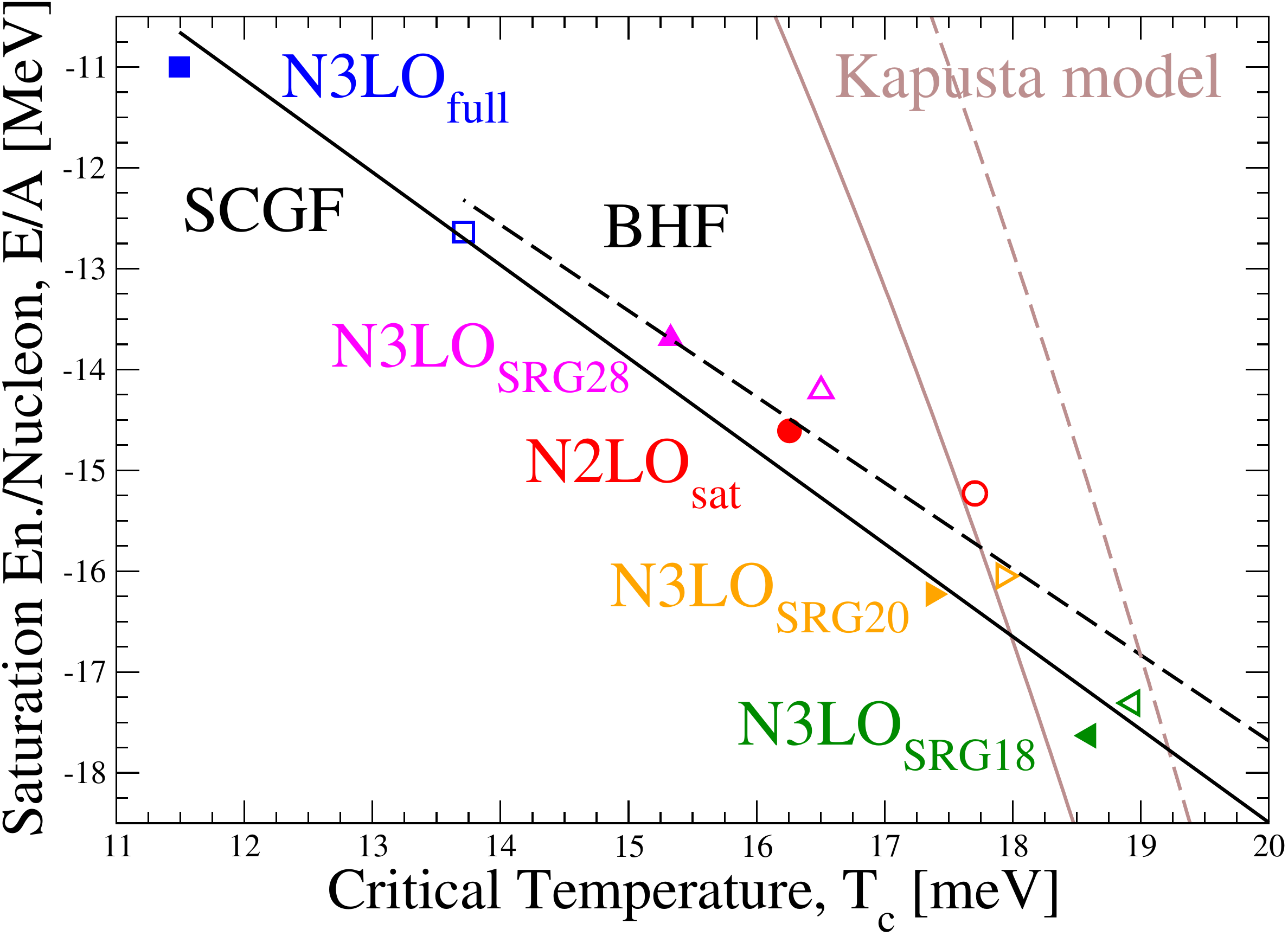}
\caption{Saturation energy per nucleon, $E_0/A$, as a function of critical temperature, $T_c$, for the chiral potentials defined in the figure. Black lines across the data points are obtained from a linear interpolation to these data. We interpolate separately for the SCGF (solid) and the BHF (dashed) cases. The steeper brown lines describe the relation arising from the Kapusta model, Eq.~(\ref{eq:kapusta_tc}), as explained in the text.}
\label{fig:tc_vs_sat}
\end{figure}

Figure~\ref{fig:tc_vs_sat} shows the zero-temperature saturation energies as a function of critical temperature for each different chiral Hamiltonian of Table~\ref{table:chiral_hamiltonian}. Details on the calculation of the saturation point will be provided elsewhere~\cite{Carboneunpub}. We present both the SCGF and the BHF results, and provide numerical data of the saturation properties in columns 4-6 of Table~\ref{table:values}. We note, in passing, that the phenomenological saturation point is not reproduced by any of our calculations, which form Coester lines above the expected values, in a similar way as presented in Ref.~\cite{Drischler2017}.  

Our results indicate that there is a clear linear correlation between saturation energy and critical temperature. We note that, while this is in agreement with the just discussed naive idea, the correlation is not easy to establish theoretically. Phenomenological models of the liquid-gas phase transition rely often on approximate expressions of the pressure, rather than the energy or the free-energy. As a consequence, the subsaturation behavior of the pressure generally dictates the predictions of $T_c$ in these phenomenological descriptions \cite{Kapusta1984,Rios2010,Rios2015}. Moreover, because the pressure is a density derivative of the energy, it does not contain any information on the actual value of the saturation energy, so it is not clear why the correlation should arise. 

Let us provide a numerical example of why this correlation is unexpected. To this end, we exploit a model by Kapusta which describes the finite-temperature behavior of a degenerate system relying on the Sommerfeld expansion~\cite{Kapusta1984}. The latter is valid at low temperatures, $T/\varepsilon_{\rm F} \ll1$, with $\varepsilon_{\rm F}$ the Fermi energy. The pressure around saturation is parametrized in terms of the compressibility, $K_0$, and the effect of in-medium correlations is summarised in an effective mass. The model predicts the following expression of the critical temperature as a function of the saturation parameters:
\begin{equation}
T_c=\frac{5^{5/6}}{2^{13/6}3^{1/3}b}\sqrt{\frac{K_0}{m_0^*}}\rho_0^{1/3}\,,
\label{eq:kapusta_tc}
\end{equation}
where $b=\big(\frac{2^{5/2}\pi}{3\hbar^3}\big)^{1/3}$.
We stress that there is no explicit dependence on the saturation energy in this formula. 

One may try to introduce a saturation energy dependence in these results by exploiting the Coester curve, relating saturation energy to saturation density in microscopic calculations \cite{Dewulf2003}. We therefore perform a linear fit to the values $(\rho_0,E_0)$ for each Hamiltonian in Table~\ref{table:values}, separately for the SCGF and the BHF results. We then insert this linear dependence, $\rho(E_0)$, into Eq.~\eqref{eq:kapusta_tc}. We set $K_0=230$ MeV, and $m_0^*=0.85m$ for SCGF data and  $m_0^*=0.76m$ for BHF data as an average of the values obtained for each Hamiltonian for the effective mass at saturation density, see Table~\ref{table:values}. This allows us to find a direct analytical relation between $T_c$ and $E_0$ which we show in Fig.~\ref{fig:tc_vs_sat}, employing both the SCGF (solid) and the BHF (dashed) results. We find that the dependence of $E_0$ on $T_c$ predicted by this variation of the Kapusta model is much steeper than what is suggested by the data. Variations on the value for the compressibility $K_0$, which for these potentials ranges between $K_0\sim[120-290]$ MeV, only shift the intercept of $T_c$ with the abscissa, but do not change the slope. Similarly, changes in values of $m_0^*$, will only change the abscissa origin but do not affect the slope. 

Before concluding this section, we stress the fact that most of the chiral forces predict a saturation point which is far from empirical expectations, see values in Table~\ref{table:values}. If we were to consider only  forces that provide a reasonable saturation point both in terms of density and energy, we would substantially reduce the uncertainty on the prediction of the critical temperature. 

In conclusion, we find a strong correlation between the saturation energy and the critical temperature. The correlation is rather linear, with a slope that is tantalizingly close to $-1$. Such a correlation is absent in density-functional calculations of the liquid-gas transition, partly because there the saturation energy is an input parameter and has little room for change \cite{Rios2010,Lourenco2016}. Even when some room is allowed, the correlation analysis of Ref.~\cite{Rios2015} indicates that $T_c$ and the saturation energy are not strongly correlated. It appears that this correlation is intrinsic to the \emph{ab initio} description of the saturation and the critical points and needs further investigation in the future.

\section{Conclusions}

We have presented a detailed study of the liquid-gas phase transition in symmetric nuclear matter, employing two different many-body schemes and five different combinations of chiral 2N and 3N forces. These five forces span a wide range of interactions, which have been renormalized in different ways and include two different orders of the chiral expansion. All our 3N forces are computed at N2LO and are included into our calculation by means of a correlated average at the one- and the two-body effective interaction levels. 

This wide set of calculations allows us to estimate two different sets of uncertainties. First, for a fixed many-body scheme, we can look at the predictions of different Hamiltonians in order to ascertain the dependence of our results on the underlying interaction. In general, we find that there is a substantial Hamiltonian dependence that, in fact, dominates across a wide range of densities and temperatures. Regarding critical properties, this Hamiltonian dependence accounts for differences in critical temperatures of the order of $5-7$ MeV. The origin of this uncertainty arises mostly from the different constraints for the LECs in the three-body part of the Hamiltonian.

Second, for a fixed Hamiltonian, we can estimate the error provided by the many-body approximation by comparing the SCGF results to those obtained within the BHF approach. We find that evolved 2N Hamiltonians generally provide SCGF and BHF results that are very close to each other (as expected). In fact, at the level of the phase transition, the differences in the many-body treatment are at most of $2$ MeV in the critical temperature. All in all, our results indicate that comprehensive calculations with a wide set of Hamiltonians and of many-body theories are necessary to provide realistic uncertainty quantification in the critical properties of nuclear matter. 

By averaging all of our results, we can estimate the critical temperature to be $T_c=16.4\pm2.3$ MeV. This average value takes into account the theoretical uncertainty coming both from the chiral Hamiltonian and from the many-body approach. Our prediction falls well within the error band provided by non-relativistic density functional calculations~\cite{Rios2010}. Furthermore, it is in good agreement with respect to recent experimentally extracted values for the nuclear matter critical temperature~\cite{Karnaukhov2009,Elliott2013}. 
Calculations could be improved in a number of ways. The inclusion of particle-hole diagrams in the calculation of the self-energy, the treatment of induced three-nucleon forces and the averaging of the 3N force are all aspects that could affect some of our conclusions at a quantitative level. We do not expect that our conclusions are changed qualitatively. 

Finally, we find an interesting and unexpected linear correlation between the saturation energy and the critical temperature provided by the different interactions. The Kapusta model, which has been used extensively in the literature to connect critical and saturation properties, cannot explain this tendency. This model relies on the Sommerfeld expansion, which is doubtfully correct across the whole density-temperature regime of the liquid-gas phase transition.

Our results represent the first nonperturbative calculation of the liquid-gas phase transition based on a wide set of modern two- and three-body chiral forces. We expect these results to be relevant for two reasons. 
On the one hand, they provide unique input towards the improvement of the construction of the chiral nuclear Hamiltonian, to predict correctly the finite-temperature properties of nuclear matter. On the other hand, our calculations pave the way towards reliable microscopic constraints to the finite-temperature equation of state of relevance in astrophysical phenomena.

\begin{acknowledgments}
This work is supported by: the STFC, through Grants No. ST/L005743/1 and ST/P005314/1;
the MICINN (Spain) through Grant No. FIS2014-54672-P from MICINN (Spain), and the Generalitat de Catalunya (Spain) through Grant No. 2014SGR-401. Partial support comes from the  Deutsche Forschungsgemeinschaft
through Grant SFB 1245 and ``PHAROS" COST Action CA16214. Calculations for this research were conducted on the Lichtenberg high performance computer of the TU Darmstadt.
\end{acknowledgments}

%\newpage %Just because of unusual number of tables stacked at end
\bibliographystyle{apsrev4-1}% Produces the bibliography via BibTeX.
\bibliography{biblio}

\end{document}